\documentclass[twocolumn,showpacs,preprintnumbers,amsmath,amssymb]{revtex4}

\usepackage{graphicx}
\usepackage{dcolumn}
\usepackage{bm}

\begin{document}

\title{New critical behavior in unconventional ferromagnetic superconductors}
\author{Dimo I. Uzunov}

\altaffiliation[]{Electronic address: uzun@issp.bas.bg. Permanent
address: CP Laboratory,
 Institute of Solid State Physics, Bulgarian Academy of Sciences, BG--1784, Sofia, Bulgaria.}
\affiliation{Max-Plank-Institut  f\"{u}r Physik komplexer Systeme,
N\"{o}tnitzer Str. 38, 01187 Dresden, Germany}

\date{2nd July 2006}
\begin{abstract} New critical behavior in unconventional
superconductors and superfluids is established and described by
the Wilson-Fisher renormalization-group method. For certain
ordering symmetries a new type of fluctuation-driven first order
phase transitions at finite and zero temperature are predicted.
The results can be applied to a wide class of ferromagnetic
superconducting and superfluid systems, in particular, to
itinerant ferromagnets as UGe$_2$ and URhGe.
\end{abstract}
\pacs{05.70.Jk,74.20.De, 75.40.Cx}
\keywords{superconductivity,
ferromagnetism, fluctuations, quantum phase transition, critical
point, order, symmetry.} \maketitle

\section{ INTRODUCTION}

In this paper new critical behavior in unconventional
ferromagnetic superconductors and superfluids is established and
described. This behavior corresponds to an isotropic ferromagnetic
order in real systems but does not belong to any known
universality class~\cite{Uzunov:1993} and, hence, it could be of
considerable experimental and theoretical interest. Due to crystal
and magnetic anisotropy, a new type of fluctuation-driven first
order phase transitions occur, as is shown in the present
investigation. The novel fluctuation effects can be observed near
finite and zero temperature (``quantum'') phase
transitions~\cite{Uzunov:1993, ShopovaPR:2003} in wide class of
ferromagnetic systems with unconventional (spin-triplet)
superconductivity or superfluidity.

The present study has been performed on the special example of
intermetallic compounds UGe$_2$ and URhGe, where the remarkable
phenomenon of coexistence of itinerant ferromagnetism and
unconventional spin-triplet superconductivity~\cite{Sigrist:1991}
has been observed~\cite{Saxena:2000}. For example, in UGe$_2$, the
coexistence phase occurs~\cite{Saxena:2000} at temperatures $0\leq
T <1$~K and pressures $1<P < P_0 \sim 1.7$~GPa. A fragment of
($P,T$) phase diagrams of itinerant ferromagnetic
compounds~\cite{Saxena:2000} is sketched in Fig.~1, where the
lines $T_F(P)$ and $T_c(P)$ of the
paramagnetic(P)-to-ferromagnetic (F) and
ferromagnetic-to-coexistence phase (C) transitions are very close
to each other and intersect at very low temperature or terminate
at the absolute zero ($P_0,0$). At low temperature, where the
phase transition lines are close enough to each other, the
interaction between the real magnetization vector
$\mbox{\boldmath$M$}(\mbox{\boldmath$r$)} =
\{M_j(\mbox{\boldmath$r$}); j=1,...,m \}$ and the complex order
parameter vector of the spin-triplet Cooper
pairing~\cite{Sigrist:1991}, $\psi(\mbox{\boldmath$r$}) =
\{\psi_{\alpha}(\mbox{\boldmath$r$}) = (\psi^{\prime}_{\alpha} +
i\psi_{\alpha}^{\prime \prime}); \alpha =1,.... n/2\}$ ($n=6$)
cannot be neglected~\cite{Uzunov:1993} and, as shown here, this
interaction produces new fluctuation phenomena.
\begin{figure}
\includegraphics[width=5.5cm,angle=-90]{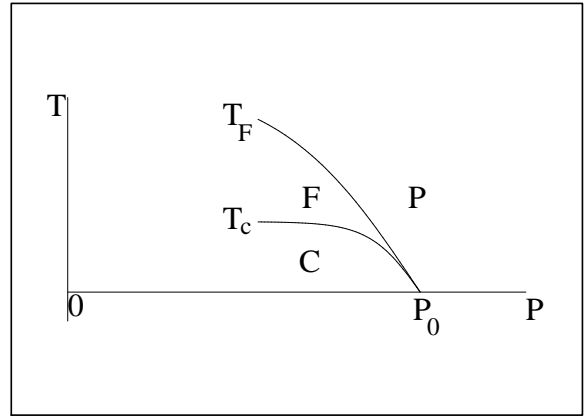}
\caption{\label{fig:UzunFF1}($P,T$) diagram with a multicritical
point $(P_0,T\sim 0)$ at very low temperature. Para- (P),
ferromagnetic (F), and coexistence (C) phases, separated by the
lines $T_f(P)$ and $T_c(P)$ of P-F and F-C phase transitions,
respectively.}
\end{figure}

Both thermal fluctuations at finite temperatures ($T>0$) and
quantum fluctuations (correlations) near the $P$--driven quantum
phase transition at $T=0$ should be considered but at the first
stage one may neglect the quantum effects~\cite{ShopovaPR:2003} as
irrelevant to finite temperature phase transitions ($T_F \sim T_c
>0$). The present treatment of recently derived free energy
functional~\cite{Machida:2001} by the standard Wilson-Fisher
renormalization group (RG)~\cite{Uzunov:1993} shows that
unconventional ferromagnetic superconductors with isotropic
magnetic order ($m=3$) exhibit a quite particular multi-critical
behavior for any $T> 0$, whereas the magnetic anisotropy ($m
=1,2$) generates fluctuation-driven first order transitions (see
Sec.~III and IV).

 As shown in Sec.~II certain terms in the general free energy
Hamiltonian~\cite{Shopova:2003}, which are relevant for the
mean-field analysis~\cite{Machida:2001, Shopova:2003} become
irrelevant within the RG framework. This leads to a considerable
simplification of the RG treatment but on the other side specific
symmetry properties of the relevant Hamiltonian terms make the
same RG treatment quite nontrivial (Sec.~III). The RG study is
performed by the known $\epsilon$-expansion (see, e.g.,
Ref.~\cite{Uzunov:1993}).

A particular feature of the present theoretical consideration is
the breakdown of usual $\epsilon$-expansion in the one-loop
approximation for RG equations $(\epsilon = 6-d)$. Thus, the
latter cannot yield conclusive results and one is forced to derive
the two-loop RG equations. The higher order of the theory restores
the $\epsilon$-expansion in a modified form - expansion in
non-integer powers ($\epsilon^{1/4}, \epsilon^{1/2},\dots$) of
$\epsilon$. In the framework of two-loop RG equations one obtains
the above mentioned new critical behavior.

This theoretical scenario looks quite similar to the breakdown of
the usual $\epsilon$-series and the
${\tilde{\epsilon}}$-expansions in non-integer powers of
${\tilde{\epsilon}}=(4-d)$, known for the first time from the
theory of critical phenomena in anisotropic disordered systems
(see, e.g., Ref.~\cite{Lawrie:1987}). It should be emphasized that
the mentioned similarity between the present theoretical analysis
and that in certain disordered systems cannot be extended beyond
some general features of the $\epsilon$-expansions. The mechanism
leading to the ${\tilde{\epsilon}}= (4-d)$-expansion in
non-integer powers of ${\tilde{\epsilon}}$ in anisotropic
disordered systems is quite different from the mechanism revealed
here. The latter is a result of the particular symmetry of the
interaction between the fields $\psi$ and $\mbox{\boldmath$M$}$.

 In order to re-derive the results in Sec.~III, one
should carefully take into account all the relevant dependencies
of the perturbation integrals on the parameters of the theory. For
this reason here the relevant perturbation contributions to the
Hamiltonian vertex parameters are presented in a general form. In
certain cases, the perturbation integrals are calculated for zero
values of the external wave vectors, or, for zero values of
certain Hamiltonian parameters~\cite{Uzunov:1993, Aharony:1974}.
In such cases some of the integrals give equal contributions to
the respective RG equations, but in other cases, for example, the
investigation of the RG stability matrix, one should carefully
take into account all the differences in the contributions of the
same integrals. The calculation of elements of the linearized RG
stability matrix is made through the exact dependence of the
perturbation terms on the parameters of the Hamiltonian and for
this aim one needs to know the perturbation integrals in their
initial general form. In all other aspects, the present
consideration follows the standard prescriptions, described in
Ref.~\cite{Aharony:1974} and applied to the study of complex
systems by means of RG and $\epsilon$-expansions in non-integer
power of $\epsilon$ (see, e.g., Ref.~\cite{Lawrie:1987}). In spite
of the specific features of the present analysis, it does not
contradict to the usual concepts and interpretations of
$\epsilon$-series (see, e.g., Ref. ~\cite{Lawrie:1987}). In this
paper some integrals appear for the first time in the RG theory of
complex systems. These integrals together with known ones are
evaluated to the accuracy needed for the RG analysis in two-loop
approximation (See Sec.~III). In RG studies in higher orders of
the loop expansion some additional relevant terms of the same
integrals should be calculated and used. In order to facilitate
the reproducibility of the results and further investigations,
some details of the calculations and an extended discussion of the
most important stages of the RG investigation are presented (Sec.
III and IV).

These remarks are important throughout the RG analysis: (1) the
calculation of fixed point (FP) coordinates, (2) the calculation
of elements of the RG stability matrix, and (3) the calculation of
critical exponents, including the stability ones. Besides, the
derivation of the RG equations should be made with a considerable
attention for reasons explained in Sec.~III. Note, that the RG
investigation can be performed in an alternative way, namely, the
RG can be applied to a new effective Hamiltonian, which is a
functional of the $\psi$-field only. This point is briefly
discussed in Sec.~II.

The consideration of quantum effects exhibits for the first time
an example of a complex quantum criticality characterized by a
double-rate quantum critical dynamics (Sec.~III.E). In the quantum
limit ($T\rightarrow 0$) the fields $\mbox{\boldmath$M$}$ and
$\psi$ have different dynamical exponents, $z_M$ and $z_{\psi}$,
and this leads to two different upper critical dimensions:
$d_U^{\psi} = 6-z_{\psi}$ and  $d_U^M = 6-z_M$. For this reason
the more complete theoretical description of quantum effects on
the properties of the zero-temperature (multi)critical point meets
difficulties (Sec.~III.E and Sec.~V). The treatment of
spin-triplet ferromagnetic superconductors with magnetic
anisotropies ($m <3$) and/or crystal symmetry lower than the cubic
one requires a somewhat different RG analysis. These systems are
considered in Sec.~IV. In Sec.~V the main results are summarized
and discussed.

\section{EFFECTIVE HAMILTONIAN}

Relevant part of the fluctuation Hamiltonian of unconventional
ferromagnetic superconductors~\cite{Machida:2001,Shopova:2003} can
be written in the form
\begin{widetext}
\begin{equation}
\label{eq1} {\cal{H}}= \sum_{\mbox{\boldmath$k$}}\left[ \left(r +
k^2 \right)|\psi(\mbox{\boldmath$k$})|^2 + \frac{1}{2}\left(t +
k^2\right)|\mbox{\boldmath$M$}(\mbox{\boldmath$k$})|^2   \right]
+\frac{ig}{\sqrt{V}}\sum_{\mbox{\boldmath$k$}_1,{\mbox{\boldmath$k$}_2}}{\mbox{\boldmath$M$}}
\left(
{\mbox{\boldmath$k$}}_1\right).\left[\psi\left({\mbox{\boldmath$k$}}_2
\right)\times \psi^{\ast}\left({\mbox{\boldmath$k$}}_1 +
{\mbox{\boldmath$k$}}_2 \right) \right]
\end{equation}
\end{widetext}
where $V \sim L^d$ is the volume of the $d-$dimensional system,
the length unit is chosen so that the wave vector
${\mbox{\boldmath$k$}}$ is confined below unity ($ 0 \leq k =
|{\mbox{\boldmath$k$}}| \leq 1)$, $g \geq 0$ is a coupling
constant, describing the effect of the scalar product of
${\mbox{\boldmath$M$}}$ and the vector product
$(\psi\times\psi^{\ast})$ for symmetry indices $n/2 = m=3$, and
the parameters $r = \alpha_s(T-T_s)$ and $t = \alpha_f (T-T_f)$
are expressed by the critical temperatures of the generic
($g\equiv 0$) superconducting ($T_s$) and ferromagnetic ($T_f$)
transitions (as usual, the parameters $\alpha_s$ and $\alpha_f$
are positive). As mean field studies
indicate~\cite{Machida:2001,Shopova:2003}, $T_s(P)$ is much lower
than $T_c(T)$ and $T_F(P) \neq T_f(P)$. As shown below the
Hamiltonian (1) describes the main fluctuation effects in a close
vicinity of critical points in spin-triplet ferromagnetic
superconductors. It is convenient to choose units, in which $k_B =
1$, and the upper cutoff for the magnitude
$k=|\mbox{\boldmath$k$}|$ of the wave vectors in Eq.~(1) is equal
to unity. Some perturbation expansions within the RG
investigation, in particular, those for isotropic systems
($n/2=m=3$) can be performed with the help of known representation
$\varepsilon_{j\alpha\beta}$ of the components $(\psi \times
\psi^{\ast})_j$ of the respective vector product in Eq.~(1).

The fourth order terms ($M^4, |\psi|^4, M^2|\psi|^2$) in the total
free energy (Hamiltonian)~\cite{Machida:2001, Shopova:2003} have
not been included in Eq.~(1) as they are irrelevant to the present
investigation. The simple dimensional analysis shows that the
$g-$term in Eq.~(1) corresponds to a scaling factor $b^{3-d/2}$
and, hence, becomes relevant below the upper borderline dimension
$d_U=6$, whereas fourth order terms are scaled by a factor
$b^{4-d}$ as in the usual $\phi^4-$theory and are relevant below
$d<4$ ($b > 1$ is a scaling number)~\cite{Uzunov:1993}. Therefore,
we should perform the RG investigation in spatial dimensions $d =
6-\epsilon$, where the $g$--term in Eq.~(1) describes the only
relevant fluctuation interaction. Besides, the total fluctuation
Hamiltonian~\cite{Machida:2001,Shopova:2003} contains off-diagonal
terms of the form $k_ik_j\psi_{\alpha}\psi^{\ast}_{\beta}$; $i\neq
j$ and/or $\alpha \neq \beta$. Using a convenient loop expansion
these terms can be completely integrated out from the partition
function to show that they modify the parameters ($r,t,g$) of the
theory but they do not affect the structure of the model (1). So,
such terms change auxiliary quantities, for example, the
coordinates of the RG fixed points (FPs) but they do not affect
the main RG results for the stability of the FPs and the values of
the critical exponents. Here these off-diagonal gradient terms
will be ignored.

The mean-field investigation of the total Hamiltonian shows that
the $g$-term in Eq.~(1) triggers the spin-triplet
superconductivity ($M$-trigger mechanism~\cite{Shopova:2003}). The
fourth order terms, mentioned above, do not play such a crucial
role. They merely stabilize the ordered phases, provide a correct
shape of the ($P,T$) phase domains for the specific crystal and
magnetic symmetries. Neglecting these terms one looses the
opportunity to describe the equilibrium order
$\langle\mbox{\boldmath$M$}\rangle = \mbox{\boldmath$M$} -
\delta\mbox{\boldmath$M$}$ and $\langle\psi\rangle = \psi -
\delta\psi $, but as shown above, the main fluctuation effects in
a close vicinity of the phase transition points can be totally
taken into account. Here the attention is focussed on the behavior
above the phase transition points where
$\langle\mbox{\boldmath$M$}\rangle = \langle\psi\rangle = 0$, and
hence,  $\mbox{\boldmath$M$} \equiv \delta\mbox{\boldmath$M$}$ and
$ \psi \equiv \delta\psi $.

The simplest theory with upper critical dimension $d_U = 6$ is
that of a classic scalar field $\varphi$ with
$\varphi^3$-interaction (see, e.g., Ref.~\cite{Uzunov:1993}).
Quite more complex problem, where $d_U=6$ and an expansions in
powers of $\epsilon=(6-d)$ are used arises in the theory of
tricritical Lifshitz points~\cite{Uzunov:1993, Uzunov:1985}. In
all these cases the upper critical dimension $d_U$ is determined
by a simple dimensional analysis in the so-called {\it tree}
approximation~\cite{Uzunov:1993} and/or by a check of
singularities of the relevant perturbation integrals (See
Sec.~III). The two mentioned examples are from theories of a
single (vector or scalar) field whereas the present theory (1)
describes two fields. Here the critical dimension $d_U =6$ is a
result of the simple fact that the total power of fields in
interaction ($g$-) term in Eq.~(1) is equal to three. Thus one may
suppose that some of the results in this paper could be applied to
critical phenomena in improper ferroelectrics~\cite{Gufan:1980},
where interactions of two fields ($\varphi_1$ and $\varphi_2$) of
type $ \varphi_1\varphi^2_2$ also occur and the upper critical
dimension is $d_U =6$.  The present investigation will
demonstrate, however, that for such interaction of two fields and
upper critical dimension $d_U=6$, the RG analysis can lead to
different results depending on the specific symmetry of the
interaction term. The outcomes are two: a lack of stable FPs and
physical arguments leading to a prediction of first order phase
transitions, or, the presence of a stable FP and, hence, a stable
critical behavior. While the particular symmetry of interaction
term provides a new stable FP and, hence, a new critical behavior
for symmetry indices $n/2=m=3$, one cannot be certain that the
same prediction can be made for ferroelectrics without a specific
investigation.

One may consider several cases: (i) isotropic systems, namely,
cubic crystal symmetry and isotropic magnetic order, when all
field components $\psi_{\alpha}$ and $M_j$ can be different zero
($n/2=m=3$), (ii) anisotropic systems when the total number $(m +
n/2)$ of field components is less than six.
 Note that the case (i) is of major
interest to real systems, where fluctuations of all field
components are possible, despite the presence of crystal and
magnetic anisotropy that nullifies some of the equilibrium field
components.

The functional integral over the fields $M_j(\mbox{\boldmath$k$})$
in the partition function $Z=\int
D\psi\mbox{\boldmath$M$}\mbox{exp}(-H/T)$ can be exactly
calculated, for example, by the total summation of perturbation
series in powers of the interaction parameter $g$. In result, one
will obtain a new effective Hamiltonian ${\tilde{H}}(\psi)$, which
is a functional of a single field - $\psi$. In this new
$\psi$-theory the effects of the magnetic ($\mbox{\boldmath$M$}$-)
subsystem will be ``hidden'' in the vertex parameters of
${\tilde{H}}(\psi)$. The effective Hamiltonian can also be treated
by RG but this task is beyond the scope of the present
consideration. The Hamiltonian (1) explicitly describes the
fluctuations of the magnetization $\mbox{\boldmath$M$}$ and this
is the advantage with respect to ${\tilde{H}}(\psi)$.
\begin{figure}
\includegraphics[width=5.5
cm,angle=-90]{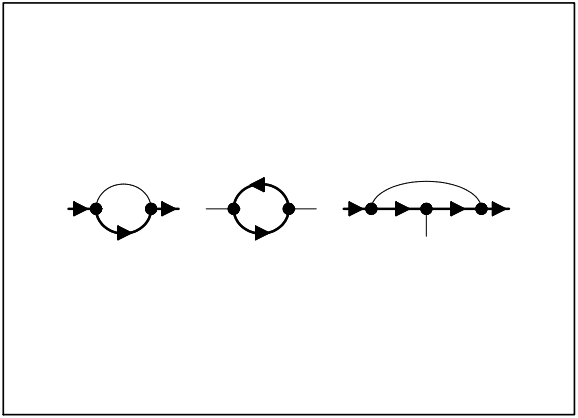} \caption{\label{fig:UzunFF2} One loop
diagrammatic contributions to the re-normalized correlation
functions $\langle |\psi_{\alpha}|^2\rangle$ and $\langle
|M_j|^2\rangle$, and to the $g$-term in Eq.~(1); ($\bullet$)
represents the $g$-interaction, the thin legs represent
$M_j(\mbox{\boldmath$k$})$, the tick legs with incoming and
outcoming arrows stand for the field components
$\psi_{\alpha}(\mbox{\boldmath$k$})$ and
$\psi_{\alpha}^{\ast}(\mbox{\boldmath$k$})$, respectively; the
internal thin and thick lines represent the bare correlation
functions $\langle |\psi_{\alpha}|^2\rangle_0$ and $\langle
|M_j|^2\rangle_0$, respectively.}
\end{figure}
\begin{figure}
\includegraphics[width=5.5cm, angle=-90]{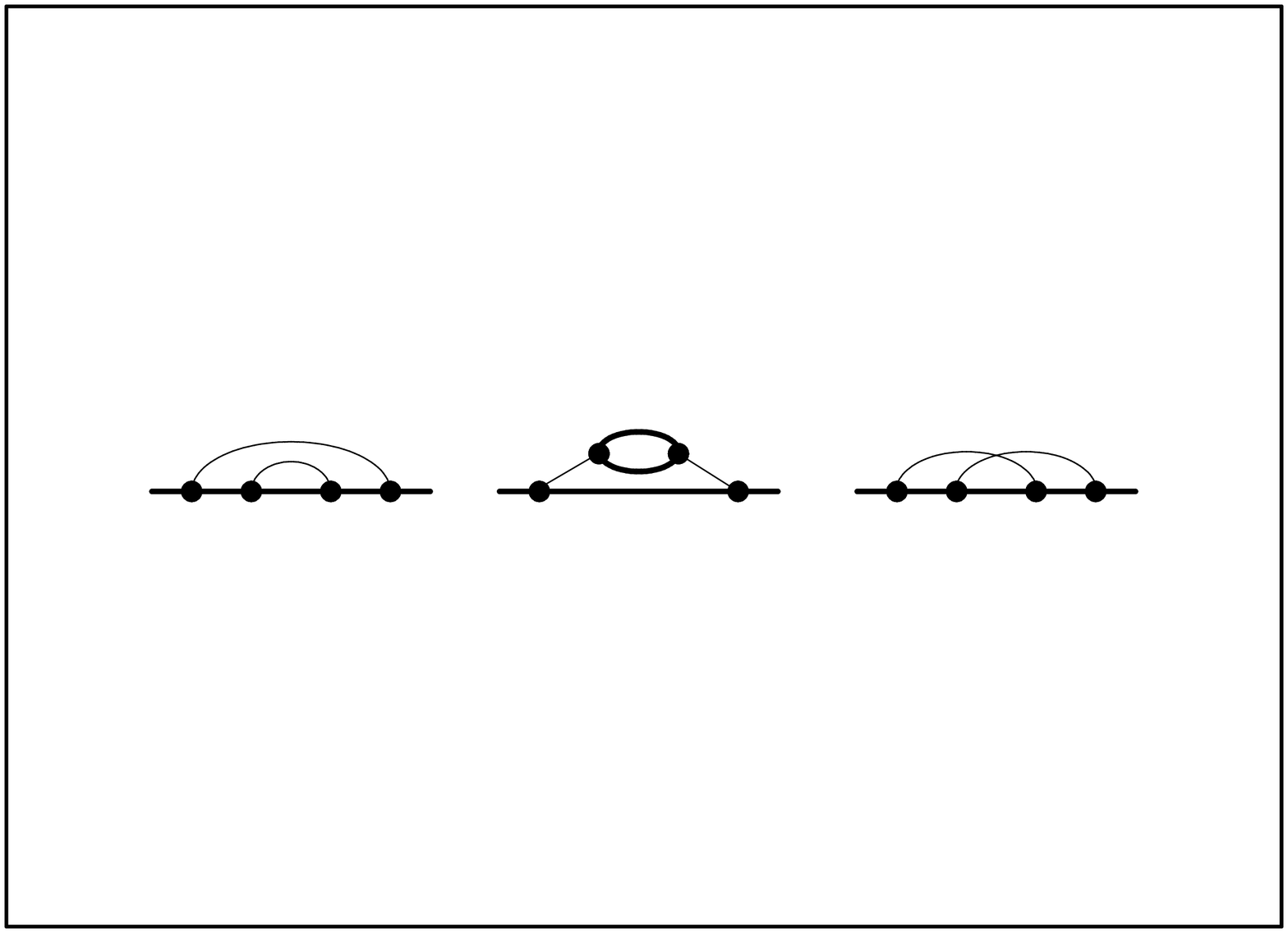}
\caption{\label{fig:UzunFF3} Two loop diagrammatic contributions
to the correlation function
$\langle|\psi_{\alpha}(\mbox{\boldmath$k$})|^2\rangle$. The
symbols have been explained in Fig.~2. The arrows of the thick
lines have been omitted.}
\end{figure}
\begin{figure}
\includegraphics[width=5.5cm, angle=-90]{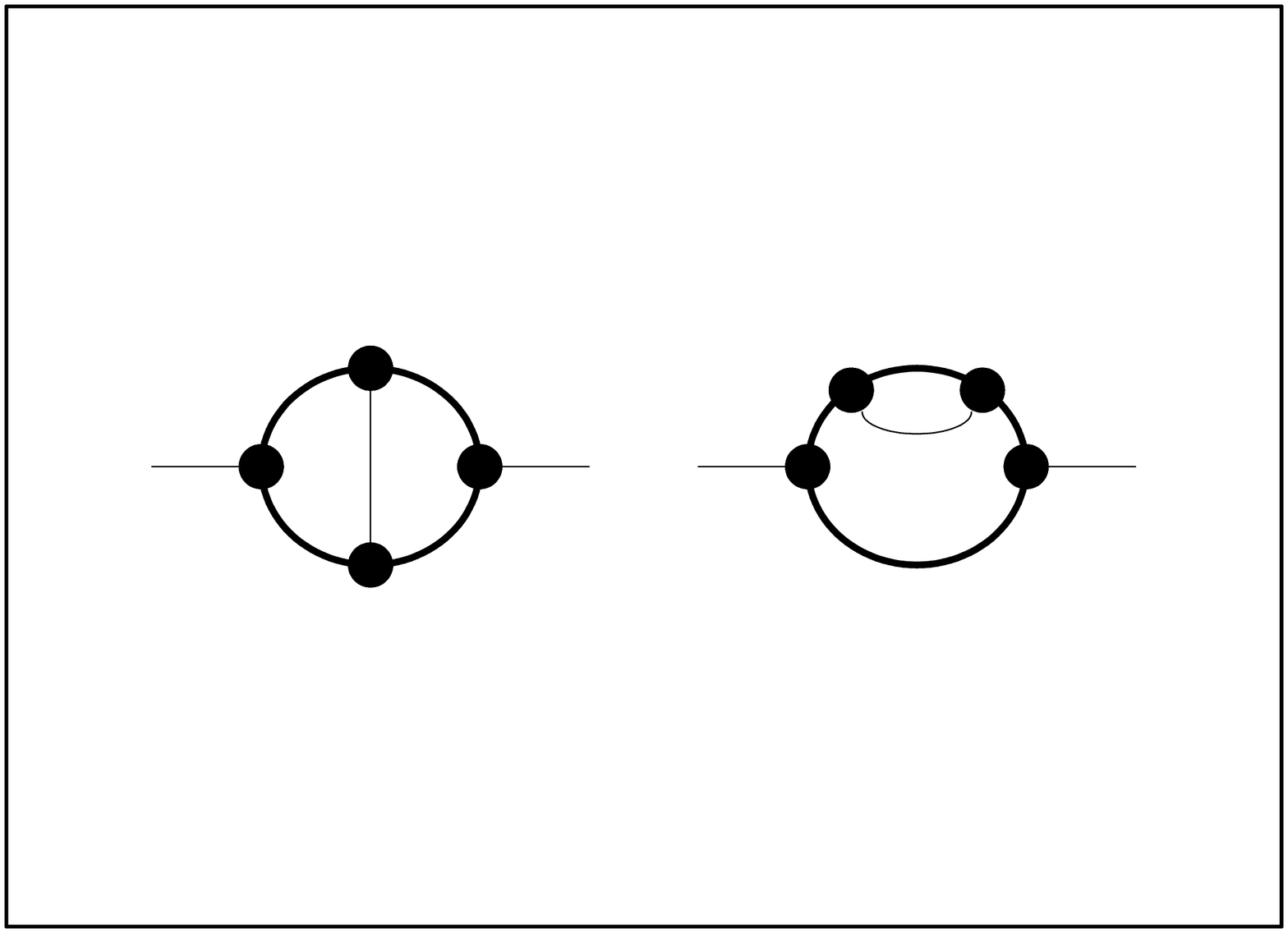}
\caption{\label{fig:UzunFF4} Two loop diagrammatic contributions
to the correlation function
$\langle|M_{j}(\mbox{\boldmath$k$})|^2\rangle$. The symbols have
been explained in Fig.~2. The arrows of the thick lines have been
omitted.}
\end{figure}
\begin{figure}
\includegraphics[width=5.2cm,angle=-90]{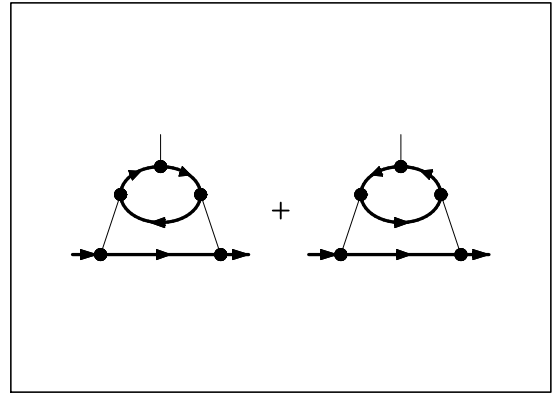}
\caption{\label{fig:UzunFF5} A sum of two loop diagrams of order
$g^5$ with a zero contribution to the renormalization of the
$g$-term in Eq.~(1). The symbols have been explained in Fig.~2.}
\end{figure}
\begin{figure}
\includegraphics[width=6cm, angle=-90]{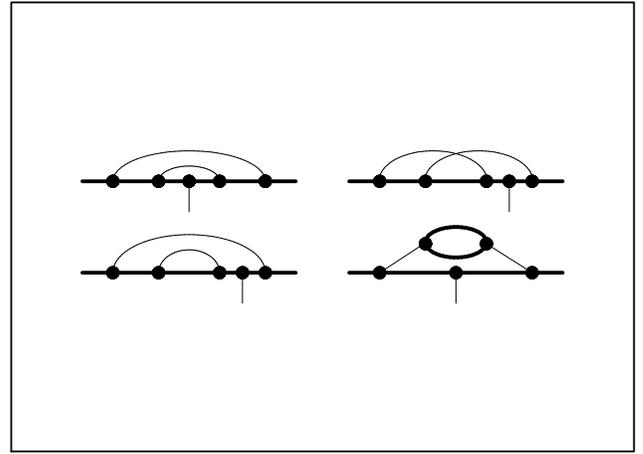}
\caption{\label{fig:UzunF3}Two loop diagrammatic contributions to
the $g$-term in Eq.~(1). The arrows of the thick lines have been
omitted.}
\end{figure}

\section{ ISOTROPIC SYSTEMS}

\subsection{RG equations}

Following Ref.~\cite{Uzunov:1993, Aharony:1974, Lawrie:1987} here
we derive the Wilson-Fisher RG equations for isotropic systems
($n/2=m=3$), described by the Hamiltonian (1), up to the two-loop
order. The main results from the RG analysis of anisotropic
systems $(n/2 + m < 6)$ can be obtained within the one-loop order
and this point will be discussed in Sec.~IV. The one loop
contributions to the RG equations are shown by diagrams in Fig.~2.
The two loop diagrammatic contributions to the renormalized
correlation functions
$\langle|\psi_{\alpha}(\mbox{\boldmath$k$})|^2\rangle$ and
$\langle|M_{j}(\mbox{\boldmath$k$})|^2\rangle$ are shown in Fig.~3
and 4, respectively.  Although the perturbation theory of the
Hamiltonian (1) is developed in a standard way, i.e., by an
expansion in powers of the interaction parameter $g$, the
derivation of the two-loop terms in the RG equations is quite
nontrivial because of the special symmetry properties of the
interaction $g$-term. For example, some diagrams with opposite
arrows of internal lines, as the couple shown in Fig.~5, have
opposite signs and compensate each other. The terms bringing
contributions to the $g$--vertex are shown diagrammatically in
Fig.~6.

In Figs.~(2) - (6) the thin external legs and thin internal lines,
corresponding to the field components $M_j(\mbox{\boldmath$k$})$
and the bare correlation function $\langle
|M_j(\mbox{\boldmath$k$})|^2\rangle_0$, respectively, can always
be supplied with incoming and/or outcoming arrows as this has been
made for the thick legs and lines, representing the fields
$\psi_{\alpha}(\mbox{\boldmath$k$})$ and
$\psi_{\alpha}^{\ast}(\mbox{\boldmath$k$})$, and the bare
correlation function $\langle
|\psi_{\alpha}(\mbox{\boldmath$k$})|^2\rangle_0$, respectively.
The thin lines in Figs.~(2) - (6) can take any orientation,
because the magnetization
$\mbox{\boldmath$M$}(\mbox{\boldmath$x$})$ is a real vector and,
hence, the Fourier amplitudes of the field components,
$M_j(\mbox{\boldmath$k$})$, obey the relation
$M^{\ast}_{j}(\mbox{\boldmath$k$})= M_j(-\mbox{\boldmath$k$})$. In
Figs.~(2) - (6) the arrows of thin lines and legs are omitted but
one should have in mind that in practical calculations with the
help of these diagrams, arrows of any orientation can be used.

The RG equations have been derived in the following general form:
\begin{widetext}
\begin{equation}
\label{eq2}
 r^{\prime} =
b^{2-\eta_{\psi}}\left\{r-2J_{sf}(r,t;0)g^2 - 2\left[2B_1(r,t,0) +
2B_2(r,t,0) + B_3(r,t,0)\right]g^4\right\}
\end{equation}
\begin{equation}
\label{eq3}
 t^{\prime} =
b^{2-\eta_{M}}\left\{t-2J_{ss}(r,r;0)g^2 - 2\left[A_1(r,t,0) +
4A_2(r,t,0)\right]g^4\right\},
\end{equation}
\begin{equation}
\label{eq4}
 g^{\prime}= b^{3-d/2-\eta_{\psi} - \eta_{M}/2}\left\{ g +
J_3(r,t)g^3 + \left[D_1(r,t) + 4D_2(r,t) +2D_3(r,t) +
2D_4(r,t)\right] g^5 \right\},
\end{equation}
\end{widetext}
\begin{equation}
\label{eq5} b^{\eta_{\psi}} = 1-2a_{sf}g^2 -(2b_1 +2b_2+b_3)g^4
\end{equation}
and
\begin{equation}
\label{eq6} b^{\eta_{M}} = 1-2a_{ss}g^2 -2(a_1 +2a_2)g^4.
\end{equation}
Here $b>1$ is a scaling number, $\eta_{\psi}$ and $\eta_{M}$ are
the Fisher exponents (anomalous dimensions of the fields $\psi$
and $\mbox{\boldmath$M$}$, respectively), and the perturbation
integrals are given by
\begin{equation}
\label{eq7}
 J_{sf}(r,t,\mbox{\boldmath$k$}) =
\int^{\prime}\frac{d^dp}{(2\pi)^d} \frac{1}{(p^2 +
t)\left[(\mbox{\boldmath$p$} + \mbox{\boldmath$k$})^2+r\right]},
\end{equation}
\begin{equation}
\label{eq8}
 J_{ss}(r,r,\mbox{\boldmath$k$}) =
J_{sf}(r,r,\mbox{\boldmath$k$}),
\end{equation}
\begin{equation}\label{eq9}
J_3(r,t) = \int^{\prime}\frac{d^dp}{(2\pi)^d} \frac{1}{(p^2 +
r)^2(p^2+t)},
\end{equation}
\begin{widetext}
\begin{equation}
\label{eq10}
 A_1(r,t,\mbox{\boldmath$k$}) =
\int^{\prime}\frac{d^dp_1d^dp_2}{(2\pi)^{2d}} \frac{1}{(p^2_1
+r)(p^2_2+r)\left[(\mbox{\boldmath$p$}_1 - \mbox{\boldmath$p$}_2
)^2+t\right]\left[(\mbox{\boldmath$p$}_1 + \mbox{\boldmath$k$})^2
+ r \right] \left[(\mbox{\boldmath$p$}_2 + \mbox{\boldmath$k$})^2
+r \right]},
\end{equation}
\begin{equation}
\label{eq11}
 A_2(r,t,\mbox{\boldmath$k$}) =
\int^{\prime}\frac{d^dp_1d^dp_2}{(2\pi)^{2d}} \frac{1}{(p^2_1
+r)^2\left[(\mbox{\boldmath$p$}_1 + \mbox{\boldmath$p$}_2
)^2+r\right]\left[(\mbox{\boldmath$p$}_1 + \mbox{\boldmath$k$})^2
+ r \right] \left[(\mbox{\boldmath$p$}_2^2 +t \right]},
\end{equation}
\begin{equation}
\label{eq12}
 B_1(r,t,\mbox{\boldmath$k$}) =
\int^{\prime}\frac{d^dp_1d^dp_2}{(2\pi)^{2d}} \frac{1}{(p^2_1
+r)^2\left[(\mbox{\boldmath$p$}_1 + \mbox{\boldmath$p$}_2
)^2+r\right]\left[(\mbox{\boldmath$p$}_1 + \mbox{\boldmath$k$})^2
+ t \right] \left[(\mbox{\boldmath$p$}_2^2 +t \right]},
\end{equation}
\begin{equation}
\label{eq13}
 B_2(r,t,\mbox{\boldmath$k$}) =
\int^{\prime}\frac{d^dp_1d^dp_2}{(2\pi)^{2d}} \frac{1}{(p^2_1
+t)^2\left[(\mbox{\boldmath$p$}_1 + \mbox{\boldmath$p$}_2
)^2+r\right]\left[(\mbox{\boldmath$p$}_1 + \mbox{\boldmath$k$})^2
+ r \right] \left[(\mbox{\boldmath$p$}_2^2 +r \right]},
\end{equation}
\begin{equation}
\label{eq14}
B_3(r,t,\mbox{\boldmath$k$}) =
\int^{\prime}\frac{d^dp_1d^dp_2}{(2\pi)^{2d}} \frac{1}{(p^2_1
+r)(p^2_2+t)\left[(\mbox{\boldmath$p$}_1 - \mbox{\boldmath$p$}_2
)^2+r\right]\left[(\mbox{\boldmath$p$}_1 + \mbox{\boldmath$k$})^2
+ t \right] \left[(\mbox{\boldmath$p$}_2 + \mbox{\boldmath$k$})^2
+r \right]},
\end{equation}
\begin{equation}
\label{eq15}
 D_1(r,t) =
\int^{\prime}\frac{d^dp_1d^dp_2}{(2\pi)^{2d}} \frac{1}{(p^2_1
+t)(p^2_2+t)(p_1^2 + r)^2\left[(\mbox{\boldmath$p$}_1 +
\mbox{\boldmath$p$}_2 )^2+r \right]^2},
\end{equation}
\begin{equation}
\label{eq16}
 D_2(r,t) =
\int^{\prime}\frac{d^dp_1d^dp_2}{(2\pi)^{2d}} \frac{1}{(p^2_1
+t)(p^2_2+t)(p_1^2 + r)^3\left[(\mbox{\boldmath$p$}_1 +
\mbox{\boldmath$p$}_2 )^2+r \right]},
\end{equation}
\begin{equation}
\label{eq17}
D_3(r,t) =
\int^{\prime}\frac{d^dp_1d^dp_2}{(2\pi)^{2d}} \frac{1}{(p^2_1
+t)^2(p^2_1+t)^2(p_2^2 + r)\left[(\mbox{\boldmath$p$}_1 +
\mbox{\boldmath$p$}_2 )^2+r \right]},
\end{equation}
\begin{equation}
\label{eq18}
D_4(r,t) =
\int^{\prime}\frac{d^dp_1d^dp_2}{(2\pi)^{2d}} \frac{1}{(p^2_1
+t)(p^2_2+t)(p_1^2 + r)^2(p_2^2 + r)\left[(\mbox{\boldmath$p$}_1 +
\mbox{\boldmath$p$}_2 )^2+r \right]},
\end{equation}
\end{widetext}
In Eqs.~(5) and (6), $a_{sf}, a_{ss},a_1,a_2,b_1,b_2$ and $b_3$
are differences of integrals, given by the rule
\begin{equation}
\label{eq19} y_l(r,t) =
\left\{\frac{1}{k^2}\left[Y_l(r,t,\mbox{\boldmath$k$}) -
Y_l,t,0)\right]\right\}_{k=0},
\end{equation}
where $Y_l$ denotes one of the integrals $J_{sf}$, $J_{ss}$, $A_j$
or $B_j$, and $y_l(r,t)$ stands for $\equiv a_l$ or $b_l$ and
accordingly, the index $l$ denotes either the symbols $sf$, $ss$
or the numbers $j= 1,\dots, 3$. In Eqs.~(7)-(18) $\int^{\prime}$
denotes an integration in restricted domains of wave
numbers~\cite{Uzunov:1993}, for example, for the integral $A_1$
this domain is given, as implied by the terms in the denominator
of the integrand, by the inequalities $1/b\leq p_i \leq 1$, $1/b
\leq|\mbox{\boldmath$p$}_1 - \mbox{\boldmath$p$}_2| \leq 1$, $1/b
\leq |\mbox{\boldmath$p$}_i + \mbox{\boldmath$k$}| \leq 1$, where
the external wave vector $\mbox{\boldmath$k$}$ obeys the condition
$0 \leq k < 1/b$, and $i= 1,2.$

The general RG Eqs.~(2)~-~(6) are the starting point of the
present RG analysis in two loop approximation. The same
Eqs.~(2)~-~(6) can be considered as reliable stage in the
derivation and analysis of RG equations in higher orders of the
loop expansion. The integrals~(7)~-~(18) correspond to the most
general diagrammatic representation
 of the respective perturbation terms given in Figs.~2-6 with the
  only difference that the dependence of the integrals
$D_j(r,t,\mbox{\boldmath$k$}_1,\mbox{\boldmath$k$}_2)$ on the
external wave vectors $\mbox{\boldmath$k$}_1$ and
$\mbox{\boldmath$k$}_2$ has been neglected as irrelevant for the
RG treatment~\cite{Uzunov:1993}, namely, $D_j(r,t)$ in
Eqs.~(15)~-~(18) denotes $D_j(r,t,0,0) $ for any $j = 1,\dots,4$.
For $d =6$ the integrals $J_3(0,0)$, $D_j(0,0)$ exhibit infrared
logarithmic divergencies at $b=\infty$, which means that the upper
critical dimension is $d_U=6$ and the $\epsilon$-expansion should
be developed in powers of $\epsilon = (6-d)$. This conclusion is
in a total conformity with the dimensional analysis, mentioned in
Sec.~II.

In the framework of the two loop approximation certain
simplifications of Eqs.~(2)-(6) can be made without any effect on
the final results of the RG analysis ~\cite{Uzunov:1993,
Aharony:1974, Lawrie:1987}. One can set $D_j(r,t) \approx D_j(0,0)
\equiv D_j $ and, hence, $D_2=D_3\equiv C_1(b)$, and
$D_1=D_4\equiv C_2(b)$, where
\begin{equation}
\label{eq20} C_1(b) =
\int^{\prime}\frac{d^dp_1d^dp_2}{(2\pi)^{2d}}
\frac{1}{p^8_1p^2_2(\mbox{\boldmath$p$}_1 + \mbox{\boldmath$p$}_2
)^2},
\end{equation}
and
\begin{equation}
\label{eq21} C_2(b) =
\int^{\prime}\frac{d^dp_1d^dp_2}{(2\pi)^{2d}}
\frac{1}{p^6_1p^4_2(\mbox{\boldmath$p$}_1 + \mbox{\boldmath$p$}_2
)^2}.
\end{equation}
Moreover, setting $r=t=0$ in Eqs.~(5) and (6) one obtains that
\begin{equation}
\label{eq22} (2b_1+2b_2+b_3) = (4a_2+a_1).
\end{equation}
 For the needs of the two loop
RG analysis, the relevant $b$-dependence of the integrals
(7)~-~(18) has been obtained in the following form:
\begin{widetext}
\begin{equation}
\label{eq23} J_{sf} = \frac{K_d}{4}\left[(2+\epsilon)(1-b^{-2}) -
4(r+t)\mbox{ln}b -2\epsilon b^{-2}\mbox{ln}b\right]
\end{equation}
\begin{equation}
\label{eq24} a_{sf} = -\frac{K_{d-1}}{32}\left[\left(2+
\frac{16\kappa\epsilon}{\pi}\right)\mbox{ln}b - (b^2-1)(r+t) +
\epsilon(\mbox{ln}b)^2\right],
\end{equation}
\end{widetext}
where
\begin{equation}
\label{eq25} \kappa = \int^{\pi}_{0}
dy\;\mbox{sin}^{4}y\left[\mbox{ln}\left(\mbox{sin}
y\right)\right]\left(4\mbox{cos}^2y-1\right),
\end{equation}
\begin{equation}
\label{eq26} a_{ss}(r,t) = a_{sf}(r,r),
\end{equation}
\begin{equation}
\label{eq27} J_3 = \frac{K_d}{2}\left[2\mbox{ln}b +
\epsilon(\mbox{ln}b)^2 - (b^2-1)(2r+t) \right]
\end{equation}
\begin{equation}
\label{eq28} a_1 = -\frac{3K^2_{d-1}}{512}\left[\mbox{ln}b +
2(\mbox{ln}b)^2\right]
\end{equation}
\begin{equation}
\label{eq29}
 a_2 =
\frac{K^2_{d-1}}{3072}\left[11\mbox{ln}b + 6(\mbox{ln}b)^2
-9(b^2-1)\right],
\end{equation}
\begin{equation}
\label{eq30} C_1 = \frac{K_{d-1}K_d}{192}\left[9(b^2-1) -
11\mbox{ln}b - 6(\mbox{ln}b)^2\right],
\end{equation}
\begin{equation}
\label{eq31} C_2 = \frac{3K_{d-1}K_d}{64}\left[\mbox{ln}b + 2
(\mbox{ln}b)^2\right],
\end{equation}
 where $K_d = 2^{1-d}\pi^{-d/2}/\Gamma(d/2)$. Note the
following useful formulae for $r=t=0$: $A_1(0,0,0) =B_3(0,0,0)$,
where
\begin{equation}
\label{eq32} A_1(0,0,0) = \frac{7}{32}K_dK_{d-1},
\end{equation}
$A_2(0,0,0)=B_1(0,0,0)=B_2(0,0,0)$, where
\begin{equation}
\label{eq33} A_2(0,0,0) = \frac{3}{32}K_dK_{d-1}\mbox{ln}b,
\end{equation}
\begin{equation}
\label{eq34} \frac{\partial A_1}{\partial r} = -4C_2,
\;\;\frac{\partial A_2}{\partial r} = -3C_1 - C_2,
\end{equation}
\begin{equation}
\label{eq35} \frac{\partial A_1}{\partial t} =
-\frac{3K_dK_{d-1}}{16}\mbox{ln}b, \;\;\;\; \frac{\partial
A_2}{\partial t} = -C_2,
\end{equation}
\begin{equation}
\label{eq36} \frac{\partial B_1}{\partial r} = -2C_1 -C_2 \;\;\;\;
\frac{\partial B_2}{\partial r} = -2C_2 - C_1,
\end{equation}
\begin{equation}
\label{eq37} \frac{\partial B_3}{\partial r} = -2C_2 -
\frac{3}{16}K_dK_{d-1}\mbox{ln}b,
\end{equation}
\begin{equation}
\label{eq38}
 \frac{\partial B_1}{\partial t} = -C_1 -C_2,
\end{equation}
and
 \begin{equation}
\label{eq39}
  \frac{\partial B_2}{\partial t} = -C_1,
\;\;\; \frac{\partial B_3}{\partial t} = -2C_2.
\end{equation}

\subsection{One loop approximation}

Neglecting the $g^4$-terms in Eqs.~(2) and (3) as well as the
$g^5$-terms in Eq.~(6),  using the respective integrals $a_{sf}$
and $s_{ss}$ as given by Eqs.~(24)-(26) to the zero approximation
in $\epsilon = (6-d)$, $r$, and $t$, we obtain that
\begin{equation} \label{eq40}
\eta_{\psi} = \eta_{M} = \frac{K_{d-1}}{8}g^{\ast 2},
\end{equation}
 where $g^{\ast}$ is any FP value of the vertex parameter $g$.
The one loop FP values $r^{\ast}$ and $t^{\ast}$ are obtained by
Eqs.~(2) and (3) where one should neglect the $g^4$-terms. The
result is
\begin{equation}\label{eq41}
 r^{\ast} = t^{\ast} =K_dg^2.
 \end{equation}
 Finally, neglecting the $g^5$-terms in
Eq.~(4) the equation for the possible FP values $g^{\ast}$ of $g$
is obtained. Taking the integral $J_3$ from Eq. (27) to zeroth
order in $\epsilon$, $r$, and $t$ one obtains that $g^{\ast}$ is
determined by the zeros of the following simple equation:
\begin{equation}\label{eq42}
g^{\ast} = b^{(\epsilon - 3\eta)/2}\left(g^{\ast}+g^{\ast 3}
K_d\mbox{ln}b\right).
\end{equation}
Using the expansion $b^{x} \approx 1 + \mbox{ln}b$, and bearing in
mind that $K_6 = 1/64\pi^3$ and $K_5= 16K_6/3$, i.e., that $K_5 =
16K_6/3$, one easily obtains that Eq.~(42) has the simple form
\begin{equation}\label{eq43}
\epsilon g^{\ast} = 0.
\end{equation}
For $\epsilon \neq 0$ this equation yields only a Gaussian FP
(GFP) ($g^{\ast} \equiv g_G = 0$, $r_G=t_G=0 $), which is stable
for dimensions $d > 6$ but is unstable for $d < 6$. As usual the
GFP describes exponents, corresponding to the Gaussian model for
$d<6$ and mean-field exponents for $d > 6$. For example, one may
use Eqs. (2)-(4) to obtain that the correlation length exponents
$\nu_{\psi}$ and $\nu_{M}$, corresponding to the correlation
lengths of the $\psi$- and $\mbox{\boldmath$M$}$-subsystems,
respectively, have no $\epsilon$-corrections $(\nu_{\psi}=\nu_{M}
= 1/2)$, and that the stability exponent, corresponding to the
parameter $g$, is given by $y_g= \epsilon/2$.

Further, Eq.~(43) shows the first particular property of the
present RG analysis which has no analog in other systems. For
$d=d_U=6$, Eq.~(43) allows an arbitrary value of $g^{\ast}$ within
the framework of usual physical conditions. This means that for
$d=6$ a new FP exists and is characterized by a real coordinate
$g^{\ast} >0$, which has no specified value and can take any
positive real value. This object can be called {\it arbitrary} FP
(in short, AFP). Note, that complex and negative real values of
$g^{\ast}$ are outside the physical domain of values of parameter
$g$. FPs with such coordinates are usually called {\it unphysical}
(UFP)~\cite{Lawrie:1987}. The exponents $\eta_{\psi}$ and
$\eta_{M}$ and the FP coordinates $r_A$ and $t_A$ are given by
Eqs. (40) and (41), where $g^{\ast} \in (0,\infty)$. The exponents
corresponding to AFP can be calculated in a standard
way~\cite{Uzunov:1993}. The results are: $\nu_1 =
(1/2-5g^2/384\pi^3)$, $\nu_2 = (1/2 + g^2/96\pi^3)$, and $y_g= 0$.
It is seen that in one loop approximation AFP has a marginal
stability $(y_g=0)$. This means that the stability properties of
this FP should be investigated in two loop order of the theory but
this is beyond the aims of this consideration. AFP exists only at
dimensions $d=6$ and, hence, it has no real physical significance.
Rather, such FPs are of pure academic interest. Thus, there is no
real motivation for its further investigation. In the remainder of
this paper only FP with some physical significance and stability
in dimensions $d < 6$ will be considered.

\subsection{Nontrivial FP in two loop approximation}

With the help of Eq.~(22), (24) -(26), (28) and (29), Eqs. (5) and
(6) can be solved with respect to the exponents $\eta_{\psi}$ and
$\eta_{M}$. One obtains that the respective equations for these
exponents are identical and, hence, the solutions should be equal
$\eta\equiv \eta_{\psi} = \eta_{M}$. In the respective equation
for the exponent $\eta$, the terms containing factors $(b^2-1)$,
see Eqs.~(24) and (29), compensate one another and only terms
containing $b$-dependent factors of type $\mbox{ln}b$ and
$(\mbox{ln}b)^2$ remain. Expanding $\eta$ to fourth order in $g$,
namely using
\begin{equation}\label{eq44}
\eta=  \alpha_1g^{\ast^2} + \alpha_2g^{\ast4},
\end{equation}
the following equation for the coefficients $\alpha_1$ and
$\alpha_2$ are obtained
\begin{widetext}
\begin{equation}\label{eq45}
\left(-\eta + \frac{K_{d-1}}{8}g^2 + \frac{\kappa \epsilon
K_{d-1}}{\pi}g^2 - \frac{13K_{d-1}^2}{768}g^4\right)\mbox{ln}b
 +
\left(\frac{\epsilon K_{d-1}}{16}g^2 -\frac{\alpha_1^2}{2}g^4
+\frac{K^2_{d-1}}{128}g^4\right)\left(\mbox{ln}b\right)^2 = 0
\end{equation}
\end{widetext}
(hereafter in this Section the superscript ($\ast$) of $g^{\ast}$
will be often omitted). The Eq.~(45) is an expansion in powers of
$g^2$ and contains terms of types: $g^2\mbox{ln}b$,
$g^4\mbox{ln}b$, $g^4(\mbox{ln}b)^2$, $\epsilon g^2\mbox{ln}b$,
and $\epsilon g^2 (\mbox{ln}b)^2$. These terms can be grouped in
three sums of terms of three types: $g^2\mbox{ln}b$,
$g^4\mbox{ln}b$, and $g^4(\mbox{ln}b)^2$. Having in mind that
Eq.~(45) should be satisfied for any $b > 1$, all sums of the
mentioned types should be equal to zero.  Taking into account
Eq.~(44) and setting the sum of terms of type $g^2\mbox{ln}b$ in
Eq.~(45) equal to zero, one easily checks the one loop result
$\alpha_1 = K_{d-1}g^{\ast 2}/8$; c.f. Eq.~(40). Now one should
set the sum of
 terms of type $g^4\mbox{ln}b$ equal to zero. This yields the
 following equation for $g$:
\begin{equation}\label{eq46}
\left(\alpha_2 + \frac{13K_{d-1}^2}{768}\right)g^2 = \frac{\kappa
K_{d-1}}{\pi}\epsilon.
\end{equation}
Now putting the sum of terms of type $(\mbox{ln}b)^2$ equal to
zero one finds
\begin{equation}\label{eq47}
\left(\alpha_1^2- \frac{K^2_{d-1}}{64}\right)g^2 =
\frac{K_{d-1}}{8}\epsilon.
\end{equation}
Eq.~(47) reproduces the one loop result for $\alpha_1$, only if
the r.h.s term is small compared to the l.h.s. ones, namely, if
$g^2 \sim \epsilon^y$, where $y < 1$. The shape of Eq.~(46) is in
a conformity with this assumption. In fact, both Eqs.~(46) and
(47) show that the opposite supposition ($y \geq 1$) does not
allow the existence of FPs of type $g^{\ast} > 0$.  Accepting the
hypothesis $y < 1$, and neglecting the r.h.s term in Eq.~(46), one
gets $\alpha_2 = -13K_{d-1}^2/768$, and, hence, Eq.~(44) takes the
form
\begin{equation}\label{eq48}
\eta=  \frac{K_{d-1}}{8}g^{2} - \frac{13K_{d-1}^2}{768}g^4,
\end{equation}
where $g \equiv g^{\ast} > 0$.

In order to find the FP values $g^{\ast}\sim \epsilon^y > 0$ of
parameter $g$, the RG Eq.~(4) should be investigated. Using the
relations between the integrals $D_j(0,0)$ and $C_1$ and $C_2$, as
discussed in Sec. III.B, this equation takes the form
\begin{equation}\label{eq49}
 g^{\prime}= b^{(\epsilon-3\eta)/2}g\left[ 1 +
J_3g^2 + 3(2C_1+C_2)g^4\right].
\end{equation}
Setting in Eq.~(49) $g^{\prime} = g \equiv g^{\ast}$ and the
values of integrals $J_3$, $C_1$ and $C_2$, given by Eqs.~(27),
(30), and (32), respectively, one obtains an equation for $g > 0$,
which contains terms with three different $b$-dependent factors:
$(b^2-1)$, $\mbox{ln}b$ and $(\mbox{ln}b)^2$. The two terms with
factors $(b^2-1)$ cancel each other, the sum of terms with
$(\mbox{ln}b)$-factors become equal to zero, as should be, if the
nonzero FP value of $g$ is given by
\begin{equation} \label{eq50}
g^{\ast}=8\left(3\pi^3\right)^{1/2}\left(2\epsilon/13\right)^{1/4}.
\end{equation}
The terms of type $g^4(\mbox{ln}b)^2$ compensate one another,
provided Eq.~(50) for the FP value of $g$ takes place. Note, that
the assumption $g^2 \sim \epsilon^y$ with $y < 1$ leads to
$\epsilon g^2 \ll g^4$ and, therefore, terms of order $\epsilon
g^2\sim \epsilon^{1+y}$ and order $\epsilon^2$ in the FP equation
for $g$ are small and can be safely ignored.

Thus the assumption $g^{\ast 2} \sim \epsilon^y$ yields an
essentially new nontrivial FP (50) for $ y = 1/2$. In the
framework of this new scheme of RG investigation the two loop
order gives RG results up to the first order in $\epsilon$ whereas
the one loop order can be used for calculation within an accuracy
of order $\epsilon^{1/2}$.

With the help of Eqs.~(2) and (3) as well as Eqs.~(8), (23), (32),
(33) and the relation between the integrals $B_j$ and $A_1$ and
$A_2$, one can easily obtain the FP coordinates
\begin{equation}\label{eq51}
r^{\ast} = K_d g^2 + \frac{g^4}{48\pi^6} + 0(g^4b^{-2}\mbox{ln}b),
\end{equation}
$r^{\ast} = t^{\ast}$, or, using Eq.~(50),
\begin{equation}\label{eq52}
r^{\ast} = t^{\ast} = 3\sqrt{\frac{2\epsilon}{13}} +
\frac{1536}{13}\epsilon.
\end{equation}

\subsection{Critical exponents in two loop approximation}

Using Eqs.~(48)and (50) the critical exponent $\eta$ can be
written in the form
\begin{equation}\label{eq53}
\eta = 2\sqrt{\frac{2\epsilon}{13}} - \frac{2}{3}\epsilon.
\end{equation}
The critical exponents $\nu_{\psi}$ and $\nu_{M}$ of the
correlation lengths of magnetic and superconduction subsystems,
respectively, as well as the stability exponent $y_{g}$ describing
the stability of the FP (50) with respect to the interaction
parameter $g$, can be obtained as eigenvalues of the linearized
stability matrix ${\hat{\mu}} =
\left(\partial\mu^{\prime}_i/\partial\mu_j\right)$ of the RG
transformation (2)~-~(4); here $\mbox{\boldmath$\mu$}
=\left(\mu_1,\mu_2,\mu_3\right) = \left(r,t,g\right)$ is a
notation of a vector in the parameter space $(r,t,g)$ of the
Hamiltonian (1). Following Refs.~\cite{Uzunov:1993, Lawrie:1987,
Aharony:1974}, the matrix elements $\mu_{11} = (\partial
r^{\prime}/\partial r)$, $\mu_{12} = (\partial r^{\prime}/\partial
t)$, $\mu_{13} = (\partial r^{\prime}/\partial g)$,\dots, are
obtained in the form:
\begin{equation}\label{eq54}
\mu_{11} = b^{2-\eta + 2x + 7x^2/3}\left\{1 + x^2\left[3(b^2-1) +
4(\mbox{ln}b)^2\right]\right\},
\end{equation}
where  $x = K_dg^2$,
\begin{equation}\label{eq55}
\mu_{12} = 2x b^{2-\eta + 4x/3}\left\{\mbox{ln}b + x\left[(b^2-1)
-\frac{2}{9}\mbox{ln}b \right]\right\},
\end{equation}
\begin{equation}\label{eq56}
\mu_{13} = \mu_{23}= -2x b^{2-\eta + 4x}\left[1 + \frac{14}{3}x
\right],
\end{equation}
\begin{equation}\label{eq57}
\mu_{21} = 4x b^{2-\eta + x}\left\{\mbox{ln}b +
\frac{x}{2}\left[3(b^2-1) -\frac{5}{3}\mbox{ln}b \right]\right\},
\end{equation}
\begin{equation}\label{eq58}
\mu_{22} =  b^{2-\eta + 4x^2}\left[1 + 4x^2(\mbox{ln}b)^2 \right],
\end{equation}
\begin{equation}\label{eq59}
\mu_{31} = 2\mu_{32} = \frac{1- b^2}{3}xg,
\end{equation}
and
\begin{equation}\label{eq60}
\mu_{33} =  b^{(\epsilon - 3\eta)/2 +x - 91x^2/36}\left\{1 +
x^2\left[(b^2-1) -\frac{8}{9}(\mbox{ln}b)^2 \right]\right\}.
\end{equation}
The solution of the eigenvalue equation $|{\hat{\mu}} -
\lambda{\hat{I}}| = 0$ is quite simple for GFP, where $\eta=x=0$.
But the solution of the same problem for the nontrivial FP, given
by Eqs. (50) and (52), is obtained by quite lengthy calculation.
Here the main steps of this calculation will be outlined. It seems
convenient to emphasize that this treatment does not depart from
the calculational schemes known from preceding
papers~\cite{Lawrie:1987, Aharony:1974}.

Having in mind that $x=K_dg^2$ and $\eta$ in Eqs.~(54)-(60) can be
represented by $\epsilon$ as given by Eq.~(50) and (53), the
coefficients of the algebraic equation of third order for
$\lambda$ are calculated as functions of $\epsilon$. The three
roots ($\lambda_1, \lambda_2, \lambda_3$) of the eigenvalue
equation are calculated in powers of $\epsilon^{1/2}$:
\begin{equation}\label{eq61}
\lambda = \beta_0 + \beta_1\epsilon^{1/2} +\beta_2\epsilon.
\end{equation}
The calculation of the eigenvalues $\lambda_j= A_j(b)b^{y_j}$ of
the matrix ${\hat{\mu}}$ should be performed very carefully, in
particular, by focussing a special attention to the behavior of
the large dangerous terms of type $b^2$ and $b^2(\mbox{ln}b)$, ($b
\gg 1$)~\cite{Aharony:1974} in the elements $\mu_{ij}$ of the
matrix $\mbox{\boldmath$\mu$}$. In the course of the calculation
the most part of these terms cancel one another and, hence, have
no effect on the final results for the critical exponents. As
usual~\cite{Aharony:1974}, some of these terms produce large
scaling amplitudes $A_j(b)$. As in the usual
$\phi^4$--theory~\cite{Aharony:1974} the amplitudes $A_j$ depend
on the scaling factor $b$. The result for the eigenvalues is
\begin{equation}\label{eq62}
\lambda_{1,2} = b^{y_{1,2}}\left(1 +
\frac{27}{13}b^2\epsilon\right),
\end{equation}
and
\begin{equation}\label{eq63}
\lambda_3 = b^{y_3}-  \frac{16}{13}(\mbox{ln}b)^2\epsilon,
\end{equation}
where the exponents
\begin{equation}\label{eq64}
y_1 =  2 + 10\sqrt{\frac{2\epsilon}{13}} + \frac{197}{39}\epsilon,
\end{equation}
\begin{equation}\label{eq65}
y_2 =  2 - 8\sqrt{\frac{2\epsilon}{13}} + \frac{197}{39}\epsilon,
\end{equation}
and
\begin{equation}\label{eq66}
y_3 =   - \epsilon,
\end{equation}
have been identified in the following way: $y_1 \equiv y_r$,
$y_2\equiv y_t$ and $y_3 \equiv y_g$. The negative sign of $y_g$
for $d < 6$ means that the nontrivial FP (50) is stable and
describes a critical behavior.

The correlation length critical exponents $\nu_{\psi} = 1/y_r$ and
$\nu_M = 1/y_t$, corresponding to the fields $\psi$ and
${\mbox{\boldmath$M$}}$, are
\begin{equation}
\label{eq67} \nu_{\psi}= \frac{1}{2} -
\frac{5}{2}\sqrt{\frac{2\epsilon}{13}} + \frac{103}{156}\epsilon,
\end{equation}
\begin{equation}
\label{eq68} \nu_M = \frac{1}{2} + 2\sqrt{\frac{2\epsilon}{13}} -
\frac{5\epsilon}{156}.
\end{equation}

These exponents describe quite particular multi-critical behavior,
which differs from the numerous examples known so far. For $d =
3$, $\nu_{\psi}$ = 0.78, which is somewhat above the usual value,
$\nu \sim 0.6 \div 0.7$ near a standard phase transition of second
order~\cite{Uzunov:1993}
 but $\nu_M = 1.76$ at the same dimension ($d=3$) is unusually large.
 The fact that the Fisher's
exponent~\cite{Uzunov:1993} $\eta$ is negative for $d=3$ does not
create troubles because such cases are known in complex systems,
for example, in conventional superconductors~\cite{Halperin:1974}.
Perhaps, a direct extrapolation of the results from the present
$\epsilon$-series is not completely reliable because of the fact
that the series has been derived under the assumptions of
$\epsilon \ll 1$ and under the conditions $\epsilon^{1/2}b < 1$,
$\epsilon^{1/2}(\mbox{ln}b) \ll 1$, provided $b > 1$. These
conditions are stronger than those in the usual
$\phi^4$-theory~\cite{Uzunov:1993,Aharony:1974}. Using the known
relation~\cite{Uzunov:1993} $\gamma = (2-\eta)\nu$, the
susceptibility exponents for $d=3$ take the values $\gamma_{\psi}
= 2.06$ and $\gamma_M = 4.65$. These values exceed even those
corresponding to the Hartree approximation~\cite{Uzunov:1993}
($\gamma = 2\nu = 2$ for $d=3$) and can be easily distinguished in
experiments.

Note, that here the interpretation of the $\epsilon$-series and,
in  particular, extrapolations of $\epsilon$-results up to
$\epsilon = 3$ should be made with a caution and according to the
remarks presented in Ref.~\cite{Lawrie:1987}. The extrapolations
of results for critical exponents, for example, Eqs.~(67) and
(68), to finite values of $\epsilon$ are
reliable~\cite{Lawrie:1987}, if and only if $\epsilon$-terms give
a small correction to the value of the respective exponent. For
example, the $\epsilon$-corrections in Eq.~(68) do not satisfy
this rule for $\epsilon = 3$ and, hence, the result for $\nu_{M}$
for $d=3$, given by Eq.~(68), seems unreliable. But the Eq.~(68)
is an useful result, which indicates that the value of the
exponent $\eta_M$ in real three dimensional systems, is large
compared to that given by usual theories and this is a reliable
conclusion. Let us emphasize that this relatively large value of
the exponent $\nu_M$ has a physical explanation in the fact the
$g$-term includes the first-order power of $\mbox{\boldmath$M$}$.
The relatively low power of $\mbox{\boldmath$M$}$ in the $g$-term
in Eq.~(1) is the reason for unusually strong magnetic fluctuation
effects on the critical behavior. These effects are responsible
for the relatively large values of the critical exponents, in
particular, for the large values of the exponents $\nu_M$ and
$\gamma_M$. This example has been given to show the way, in which
the present results should be interpreted in discussion of real
three dimensional systems.

\subsection{Notes about the quantum criticality at zero temperature}

The critical behavior discussed so far may occur in a close
vicinity of finite temperature multi-critical points ($T_c=T_f>0$)
in systems possessing the symmetry of the model (1). In certain
systems, as shown in Fig.~1, this multi-critical points may occur
at $T=0$. In the quantum limit ($T\rightarrow 0$), or, more
generally, in the low-temperature limit [$T \ll \mu; \mu\equiv
(t,r);k_B=1$] the thermal wavelengths of the fields
$\mbox{\boldmath$M$}$ and $\psi$ exceed the inter-particle
interaction radius and the quantum fluctuations become essential
for the critical behavior~\cite{ShopovaPR:2003,Hertz:1976}.

The quantum effects can be considered by RG analysis of
comprehensively generalized version of the model~(1), namely, the
action ${\cal{S}}$ of the referent quantum system. The generalized
action is constructed with the help of the substitution
$(-{\cal{H}}/T) \rightarrow S[{\mbox{\boldmath$M$}}(q),\psi(q)]$.
The description will be given in terms of the (Bose) quantum
fields $\mbox{\boldmath$M$}(q)$ and $\psi(q)$, which depend on the
$(d+1)$-dimensional vector $q = (\omega_l,
{\mbox{\boldmath$k$}})$; $\omega_l = 2\pi lT$ is the Matsubara
frequency ($\hbar=1;l = 0, \pm1,\dots$). The
${\mbox{\boldmath$k$}}$-sums in Eq.~(1) should be substituted by
respective $q$-sums and the inverse bare correlation functions ($r
+ k^2$) and ($t + k^2$) in Eq.~(1) contain additional
$\omega_l-$dependent terms, for example~\cite{ShopovaPR:2003,
Hertz:1976}
\begin{equation} \label{eq69}
\langle|\psi_{\alpha}(q)|^2\rangle^{-1} = |\omega_l|+ k^2 + r.
\end{equation}
The bare correlation function $\langle|M_j(q)|^2\rangle$ contains
a term of type $|\omega_l|/k^{\theta}$, where $\theta = 1$ and
$\theta =2$ for clean and dirty itinerant ferromagnets,
respectively~\cite{Hertz:1976}.

Quantum dynamics of the field $\psi$ is described by the bare
value $z=2$ of dynamical critical exponent $z=z_{\psi}$ whereas
the quantum dynamics of the magnetization corresponds to $z_M = 3$
(for $\theta =1$), or, to $z_M = 4$ (for $\theta = 2$). This means
that the classical-to-quantum dimensional crossover at
$T\rightarrow 0$ is given by $d \rightarrow (d + 2)$ and, hence,
the system exhibits a simple mean field behavior for $d \geq 4$.
Just below the upper quantum critical dimension $d_U^{(0)} =4$ the
relevant quantum effects at $T=0$ are represented by the field
$\psi$ whereas the quantum $(\omega_l$--) fluctuations of
magnetization are relevant for $d < 3$ (clean systems), or, even
for $d < 2$ (dirty limit)~\cite{Hertz:1976}. This picture is
confirmed by the analysis of singularities of relevant
perturbation integrals. Therefore, the quantum fluctuations of the
field $\psi$ have a dominating role for spatial dimensions $d <4$.

Taking into account the quantum fluctuations of the field $\psi$
and completely neglecting the $\omega_l$--dependence of the
magnetization ${\mbox{\boldmath$M$}}$, $\epsilon_0 =
(4-d)$--analysis of the generalized action ${\cal{S}}$ has been
performed within the one-loop approximation (order
$\epsilon_0^1$). In the classical limit ($r/T \ll 1$) one
re-derives the results, already reported above together with an
essentially new result, namely, the value of the dynamical
exponent $z_{\psi}= 2 - (2\epsilon/13)^{1/2}$, which describes the
quantum dynamics of the field $\psi$. In the quantum limit ($r/T
\gg 1$, $T\rightarrow 0$), static phase transition properties are
affected by the quantum fluctuations, in particular, in isotropic
systems ($n/2=m=3$). In this case, the one-loop RG equations,
corresponding to $T = 0$, are not degenerate and give definite
results. The RG equation for $g$,
\begin{equation} \label{eq70}
g^{\prime}=
b^{\epsilon_0/2}g\left(1 + \frac{g^2}{24\pi^3}\mbox{ln}b\right),
\end{equation}
yields two FPs: ({\em a}) a Gaussian FP ($g_G =0$), which is
unstable for $d<4$, and ({\em b}) a FP $(g^2)^{\ast} =
-12\pi^3\epsilon_0$, which is unphysical [$(g^2)^{\ast}<0$]  for
$d<4$ and unstable for $d \geq 4$. Thus the new stable critical
behavior, corresponding to $T>0$ and $d<6$, disappears in the
quantum limit $T\rightarrow 0$.

At absolute zero ($T=0$) and any dimension $d > 0$ the $P-$driven
phase transition (Fig.~1) is of first order. This can be explained
as a mere result of the limit $T \rightarrow 0$. The only role of
the quantum effects is the creation of the new unphysical FP ({\em
b}). In fact, the referent classical system described by
${\cal{H}}$ from Eq.~(1) also looses its stable FP (8) in the
zero-temperature ({\em classical}) limit $T\rightarrow 0$ but does
not generate any new FP, because of the lack of $g^3$-term in the
equation for $g^{\prime}$; see Eq.~(13). At $T=0$ the classical
system has purely mean field behavior~\cite{ShopovaPR:2003}, which
is characterized by a Gaussian FP ($g^{\ast} = 0$) and is unstable
towards $T$--perturbations for $0<d<6$. This is a usual classical
zero temperature behavior, where the quantum correlations are
ignored. For the standard $\phi^4-$ theory this picture holds true
for $d<4$.

One may suppose that the quantum fluctuations of the field $\psi$
are not sufficient to ensure a stable quantum multi-critical
behavior at $T_c=T_F=0$ and that the lack of such behavior is a
result of neglecting the quantum fluctuations of
$\mbox{\boldmath$M$}$. One may try to take into account these
quantum fluctuations by quite special approaches from the theory
of disordered systems, where additional expansion parameters are
used to ensure the marginality of the fluctuating modes at the
same borderline dimension $d_U$. It may be conjectured that the
techniques, known from the theory of disordered systems with
extended impurities, cannot be straightforwardly applied to the
present problem and, perhaps, a completely new approach should be
introduced.

\section{ ANISOTROPIC SYSTEMS}

For anisotropic systems, mentioned in Sec.~II as case (ii), the RG
analysis leads to different results. The RG equations have no
stable FP for dimensions $d < 6$ and, therefore, one may conclude,
quite reliably, that the
$M_j\psi_{\alpha}\psi_{\beta}$-interactions in the Hamiltonian (1)
induce fluctuation-driven first order phase transitions. The RG
result for a lack of FP is not enough to make conclusions about
the order of the phase transition, but in our case there are
additional strong heuristic arguments as well as arguments from
preceding mean-field investigations~\cite{Shopova:2003}. The
heuristic argument is that the account of anisotropy effects often
leads to change of the phase transition order from second order
phase transition, when such effects are ignored to a first order
phase transition, when the anisotropy is properly taken into
account (see, e.g., Ref.~\cite{Uzunov:1993}). Besides, a recent
mean field study shows that first order phase transitions and
multi-critical points are likely to occur in both isotropic and
anisotropic ferromagnetic superconductors for broad variations of
the theory parameters $(r,t,g,...)$. These non-RG arguments
strongly support the present point of view that the lack of FP of
RG equations for anisotropic systems can reliably be interpreted
as indication for fluctuation-driven first order phase transitions
in dimensions $d < 6$ and, in particular, for $d =3$.

Now, one has to prove that RG equations, corresponding to
anisotropic systems do not exhibit stable FP for $d < 6$. The case
(ii) of anisotropic systems, mentioned in Sec.~II, contains
several subclasses of systems. For example, one may have cubic
crystal anisotropy with $XY$ magnetic symmetry ($M_1,M_2,0$),
i.e., magnetic symmetry index $m =2$, or, magnetic anisotropy of
Ising type: $(0,0,M_3)$, i.e., $m =1$. Another example is the
tetragonal crystal anisotropy, where the symmetry index for the
superconducting order is $n/2 =2$, i.e., $(\psi_1,\psi_2,0)$ and
the magnetic symmetry can be of two types:  XY order symmetry ($m
=2$) and Ising symmetry ($m=1$). These and other examples of
anisotropic systems with total symmetry index $(n/2 + m) < 6$ can
be considered separately to prove the lack of stable FP in each of
the cases.

We shall demonstrate the lack of stable FP for $d < 6$ for
uniaxial (Ising) magnetic anisotropy $(0,0,M_3)$  and cubic
crystal symmetry ($n/2=3$). In this case, the one loop RG
equations have the form
\begin{equation} \label{eq71}
r^{\prime} = b^{2-\eta}\left[r - J_{sf}(r,t,0)\right],
\end{equation}
\begin{equation} \label{eq72}
t^{\prime} = b^{2-\eta_{3}^{\prime}}\left[r -
2J_{ss}(r,r,0)\right],
\end{equation}
and
\begin{equation} \label{eq73}
 g^{\prime} = b^{3-(d +\eta_1 + \eta_2 + \eta_M)/2}\left[g + J_3(r,t)g^3
 \right],
\end{equation}
where the exponents $\eta_1 \;(= \eta)$, $\eta_2\; (=\eta)$ and
$\eta_M$ correspond to the fields components $\psi_1$, $\psi_2$,
and $M_3$, respectively. The exponents $\eta_1$ and $\eta_2$ are
equal because the equations for them are the same
\begin{equation}
\label{eq74} b^{\eta_{1,2}} = 1- a_{sf}g^2.
\end{equation}
From Eq.~(74) one obtains $\eta_1=\eta_2\equiv \eta =
K_{d-1}g^2/16$. Because of this equality of $\eta_1$ and $\eta_2$,
 Eq.~(71), which describes the RG transformations of the two
Hamiltonian terms of type $r|\psi_1|^2$ and $r|\psi_2|^2$,  is the
same. This fact ensures a self-consistency of RG transformation
for the parameter $r$. The one loop equation for the exponent
$\eta_M$ is given by Eq.~(7), provided  the $g^4$-term in the
r.h.s. of this equation is ignored. Thus one finds that $\eta_M =
2\eta= K_{d-1}g^2/8$ as is in isotropic systems.

It should be emphasized that the perturbation series does not give
self energy contributions to the $|\psi_3|^2$-term. That is why
the RG transformation is performed after a simple procedure of
integration out of this field in the functional integral for the
system partition function. The integration is exact as the
respective integral is Gaussian. The effect is an irrelevant
contribution to the system free energy. Thus, the field component
$\psi_3$ is redundant in this consideration and does not
participate in the RG transformation. On the other hand, if one
keeps the field component $\psi$ in this RG transformation and
considers it at the same footing as the other fields $(\psi_1,
\psi_2, M_3)$ the resulting RG transformation of the term $(r +
k^2)|\psi_3(k)|^2$ in Eq.~(1) will produce a transformation for
the parameter $r$, $r^{\prime} = b^2r$, which contradicts to the
relation (71). Hopefully, this is not the right way of RG
treatment and here its discussion should be understood only as a
note of caution.

Using Eq.~(73), Eq.~(27) for the integral $J_3$ to zeroth order in
$\epsilon$, $r$, and $t$, as well as the above results for the
exponents $\eta$ and $\eta_m$, one can easily obtain FPs. The GFP
($g_G = 0$) exists and is stable for $d
>6$. The second FP is given by $g^{\ast 2} = - 96 \pi^3\epsilon$,
which means that this FP is unphysical for $\epsilon >0$, i.e.,
for dimensions $d<6$. For $d > 6$ this FP is physical, i.e.,
$g^{\ast}$ has a real positive value but for these high dimensions
it is unstable towards the parameter $g$ ($y_g = -11\epsilon/2 >
0$ for $\epsilon < 0$). In this way we proved the lack of stable
FP for dimensions $d< 6$. The result cannot be changed in the next
orders of the loop expansion.

The RG analysis, performed above, is valid also for systems with
uniaxial magnetic anisotropy ($0,0,M_3$) and tetragonal crystal
symmetry, $(\psi_1, \psi_2,0)$. The only difference is that in
these systems the redundant field component $\psi_3$ is equal to
zero and one does not need to perform a functional integration
over redundant fields.

For tetragonal symmetry and bi-axial (XY) magnetic anisotropy
$(M_1, M_2,0)$, the $g-$term in Eq.~(1) is equal to zero and,
hence, Eq.~(1) describes a simple Gaussian fluctuations. In this
case one must consider the fluctuation effects coming from the
fourth order terms in the general effective
Hamiltonian~\cite{Machida:2001, Shopova:2003}. These terms may
lead to the appearance of familiar types of FPs, which describe
bicritical and tertracritical points of phase transitions (see,
e.g., Ref.~\cite{Uzunov:1993}).

A reliable conclusion for the fluctuation-driven first order phase
transition in systems with cubic crystal symmetry and XY magnetic
order $(M_3\equiv 0)$ can be made only after a check by the one
loop RG investigation, but the results established so far imply
that the appearance of stable FP in such systems is quite
unlikely. The common difference in the results from one loop RG
equations for isotropic and anisotropic unconventional
ferromagnetic superconductors is in number factors and scaling
exponents of type $b^{\eta}$ in the respective RG equations. The
new stable FP (50) occurs as a result of quite special set of
number coefficients and scaling factors in the RG equations, which
is not the case in anisotropic systems; c.f. the one loop order
terms in Eqs.~(2)-(6) on one hand and the terms in Eqs.~(71)-(74)
on the other hand.

\section{CONCLUSION}

The general RG equations for ferromagnetic superconductors with
spin-triplet Cooper pairing were derived and analyzed up to the
second order in the loop expansion.  For cubic crystals with
isotropic magnetic order a new universality class of critical
behavior was predicted. The main features of this new critical
behavior were established and the critical exponents were
calculated. It has been shown that lower crystal and magnetic
symmetry produces different fluctuation effect - a fluctuation
change of the phase transition order. The way of interpretation of
the results and the extrapolation of $\epsilon$-series to real
dimensions ($d=3$) have been discussed at the end of Sec.~III.D.
The strong fluctuation interactions of type
$M_j\psi_{\alpha}\psi_{\beta}$ have a crucial effect on the
quantum criticality at zero temperature and some new features of
this quantum phase transitions have been outlined in Sec.III.E.
The complete understanding of this quantum phase transition
requires further theoretical investigations and, perhaps, some new
ideas of calculation. As mentioned in Sec.~III.E, the satisfactory
consideration of the quantum fluctuations of both fields
$\mbox{\boldmath$M$}$ and $\psi$ requires a new RG approach, in
which one should either consider the difference $(z_M-z_{\psi})$
as an auxiliary small parameter or invent a completely new
theoretical scheme of description. This problem is quite general
and presents a challenge to the theory of quantum phase
transitions~\cite{ShopovaPR:2003}. Within this research the
present author has not been able to give a comprehensive solution
of the problem and, hence, the discussion of quantum effects
presented in Sec.~III.E should be accepted as a preliminary
outline of general problems, rather than a report of a complete
description of this type of quite complex quantum phase
transitions.

The results can be of use in interpretations of recent
experiments~\cite{Pfleiderer:2002} in UGe$_2$, where the magnetic
order is uniaxial (Ising symmetry) and the experimental data, in
accord with the present consideration, indicate that the C-P phase
transition is of first order. Systems with isotropic magnetic
order are needed for an experimental test of the newly described
multicritical behavior. The present results can be applied to any
natural system of the same class of symmetry, although this report
is based on particular example of itinerant ferromagnetic
compounds.

\vspace{0.5cm} {\bf \small ACKNOWLEDGEMENTS}

The author thanks for the hospitality of JINR-Dubna,
MPI-PKS-Dresden, and ICTP-Triest where parts of this research have
been carried out. Financial support by grants No. P1507 (NFSR,
Sofia) and No. G5RT-CT-2002-05077 (EC, SCENET-2, Parma) is also
acknowledged.

\end{document}